\documentclass{webofc}
\usepackage[varg]{txfonts}   % Web of Conferences font

\usepackage{booktabs}
\usepackage{slashbox}
\begin{document}
\title{Using machine learning to speed up new and upgrade detector studies: a calorimeter case}
%
% subtitle is optional

\author{\firstname{Fedor} \lastname{Ratnikov}\inst{1,2}\fnsep\thanks{\email{fedor.ratnikov@gmail.com}} \and
        \firstname{Denis} \lastname{Derkach}\inst{1} \and
        \firstname{Alexey} \lastname{Boldyrev}\inst{1}\fnsep\thanks{\email{alexey.boldyrev@cern.ch}} \and
        \firstname{Andrey} \lastname{Shevelev}\inst{1} \and
        \firstname{Pavel} \lastname{Fakanov}\inst{1} \and
        \firstname{Leonid} \lastname{Matyushin}\inst{1}
        % etc.
}

\institute{National Research University Higher School of Economics,\\Laboratory of Methods for Big Data Analysis, 11 Pokrovsky blvd., Moscow 109028, Russia
\and
           The Yandex School of Data Analysis, 11/2 Timura Frunze St., Moscow 119021, Russia
          }

\abstract{%
% Designing new experiments, as well as upgrading of ongoing experiments, is a continuous process in experimental high energy physics. Frontier R\&Ds are used to squeeze the maximum physics performance using cutting edge detector technologies.\par

% The evaluation of physics performance for a particular detector configuration includes sketching this configuration in \textsc{Geant4}\cite{geant4}, simulating typical signals and backgrounds, reconstructing the attributes of physical objects and combining results into final quality metrics. Since the best solution is always a trade-off between different kinds of limitations, a quick turn over is necessary to evaluate physics performance for different technical solutions in different detector configurations.\par

% Two typical problems which slow down evaluating physics performance for different detector technologies and configurations are: describing \textsc{Geant4} geometry together with a signal processing chain for an adequate description of the detector response and developing an adequate algorithm for physics reconstruction of the detector response under study. Both problems may be addressed using ML (machine learning) approaches.\par
% % In addition to this, the whole procedure can be viewed as a black-box optimisation, which gives access to numerous available methods.\par

In this paper, we discuss the way advanced machine learning techniques allow physicists to perform in-depth studies of the realistic operating modes of the detectors during the stage of their design. Proposed approach can be applied to both design concept (CDR) and technical design (TDR) phases of future detectors and existing detectors if upgraded. The machine learning approaches may speed up the verification of the possible detector configurations and will automate the entire detector R\&D, which is often accompanied by a large number of scattered studies. We present the approach of using machine learning for detector R\&D and its optimisation cycle with an emphasis on the project of the electromagnetic calorimeter upgrade for the LHCb detector\cite{lhcls3}. The spatial reconstruction and time of arrival properties for the electromagnetic calorimeter were demonstrated.
% To obtain the planned physical performance during the R\&D of modern experiments in HEP, the detailed \textsc{Geant4} simulations are necessary. Such simulations are computationally expensive when 

% to speed up the detector development and optimisation cycle with an emphasis on the project of the calorimeter upgrade for the LHCb detector\cite{lhcls3}.

% The calorimeters are an essential part of most of the existing and developing detectors in high energy physics. The high luminosity of the collider causes a high multiplicity and hit occupancy in the calorimeter.

% The methods presented in this paper, allow physicists to perform in-depth studies for the realistic operating modes of the calorimeters during the stage of their design.

% Proposed calorimeter optimisation approach can be applied to both design concept (CDR) and technical design (TDR) phases of both future calorimeter detectors and existing calorimeters if upgraded. This approach will speed up the verification of calorimeter configurations and will automate the entire calorimeter R\&D, which is often accompanied by a large number of scattered studies.
}
\maketitle
\section{Introduction} \label{intro}
The calorimeters are an essential part of most of the existing and developing detectors in high energy physics. The high luminosity delivered by the collider causes a high multiplicity and hit occupancy in the calorimeter. In order to operate under such conditions a new generation of the calorimeters is characterised by high granularity (increased number of channels) and by the ability to measure the time of arrival of the particles to mitigate pile-up.\par

To obtain the planned physical performance during the R\&D of modern experiments in HEP, the detailed \textsc{Geant4} simulation\cite{geant4} of the calorimeter is necessary. Such simulations are computationally expensive taking into account the large number of channels and the variety of possible options in the calorimeter module technologies, in the modules arrangement, in the reconstruction of the attributes of physical objects, etc. The optimisation cycle, within the calorimeter R\&D, comprises several computationally intensive elements, such as shower development and particle transport. Processes of multi-parametric optimisation appear to be also expensive. These factors make new approaches to calorimeter development necessary. Machine learning allows a quick turnover for the optimisation cycle, when parameters are changed, and eliminates manual work for re-tuning the simulation and reconstruction.

% Reproducing a high multiplicity of primary interactions is difficult in experiments on test beams, which are a necessary step in calorimeter R\&D. The significantly increased number of channels leading to high granularity makes full \textsc{Geant4}\cite{geant4} simulation of a calorimeter response a computationally expensive task.

% Generative model trained on the sample produced by
% \textsc{Geant4} for LHCb-inspired shashlik ECAL technology
% produces good enough clusters at $10^{-5}$ smaller CPU time.\\

% \section{optimisation cycle involving Machine Learning}

\section{Spatial Reconstruction}
To reconstruct the hit position of the particle reached calorimeter, we implement an approach which is based on \textsc{Pythia8}-generated events of $B_s^0 \rightarrow J/\psi (\rightarrow \mu^{+}\mu^{-}) \pi^0 (\rightarrow \gamma \gamma)$ (hereinafter \textit{signal sample}), generated with default LHCb tunings and on a data of \textsc{Geant4}-simulated events in the simplified high-granularity detector.
This simplified simulation setup uses the same alternating layers of scintillator and lead plates (\textit{Shashlik} technology), as in LHCb Electromagnetic calorimeter (ECAL)\cite{lhcbmain} and it consists of a matrix of 30x30 cells of size 20.2x20.2~mm$^2$ in $\eta$ -- $\phi$ plane.
This allows us to emulate each type of current ECAL modules: inner, middle or outer with cells size of 40.4x40.4~mm$^2$, 60.6x60.6~mm$^2$ or 121.2x121.2~mm$^2$, respectively.

\begin{figure} [hbt]
\centering
\includegraphics[width=0.9\linewidth]{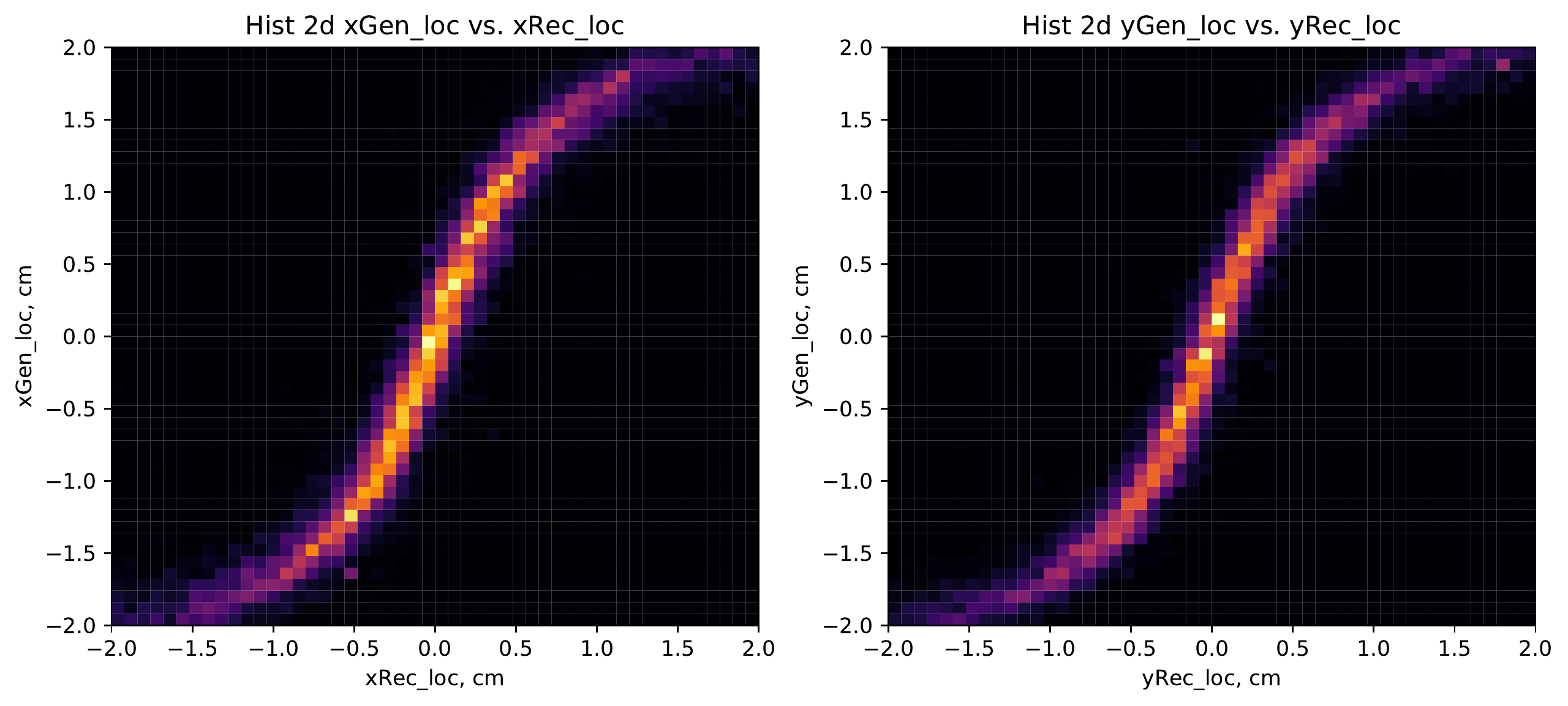}
\caption{\label{fig:inner_0_nPV_s_curve_x_and_y} The dependence (2-dimensional distribution) of local coordinates of signal cluster barycentre on true coordinates of the hit position for inner modules. The local coordinates are of \textit{x} (left) and \textit{y} (right). Here and in the Figure~\ref{fig:S-curve_calibration} colour from violet (dark) to yellow (bright) represents the normalised counts of the events from 0.0 to 1.0, respectively.}
\end{figure}

\begin{figure} [hbt]
\centering
    \begin{minipage}{.45\textwidth}
        \centering
        \includegraphics[width=\linewidth]{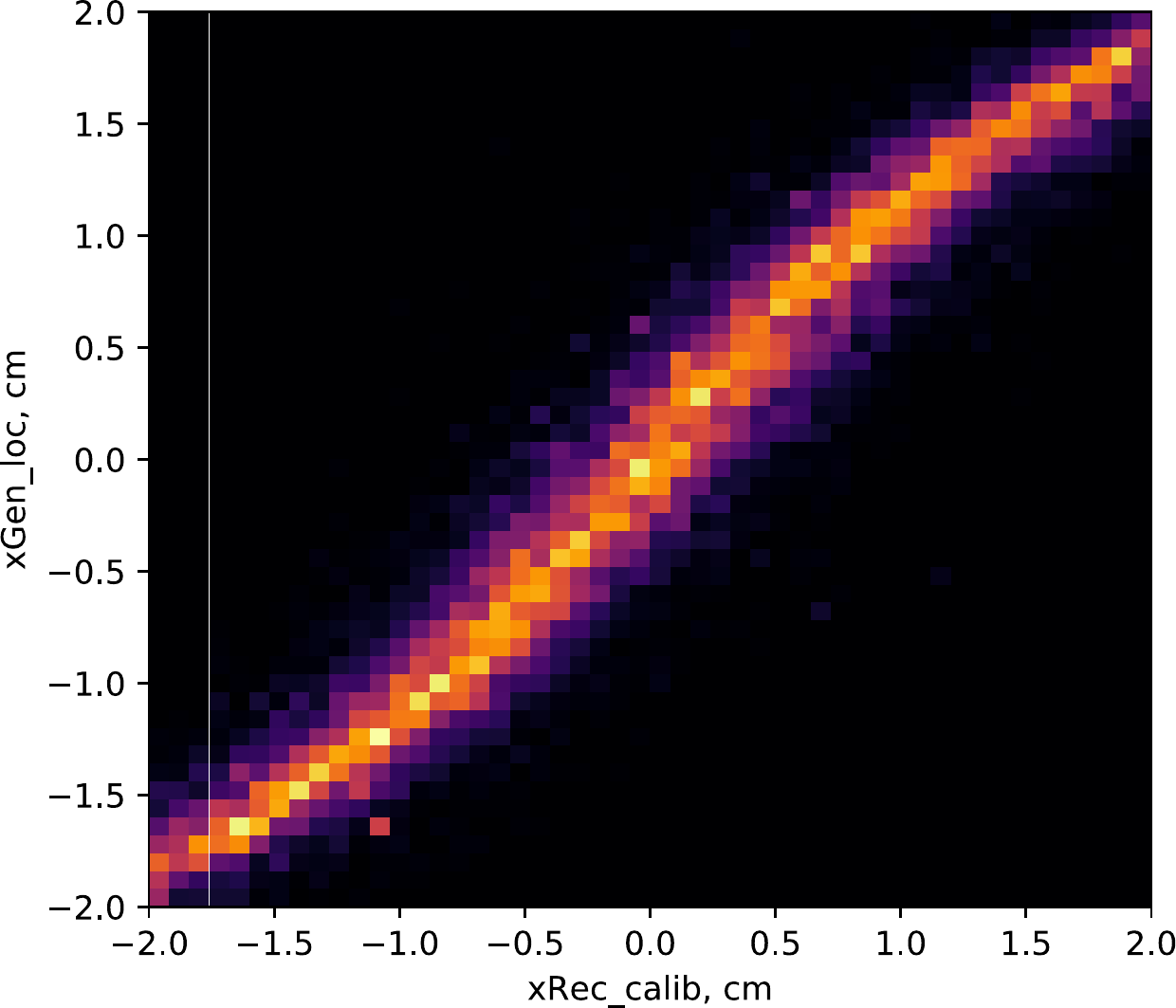}
    \end{minipage}%
\centering
    \begin{minipage}{.45\textwidth}
        \centering
        \includegraphics[width=\linewidth]{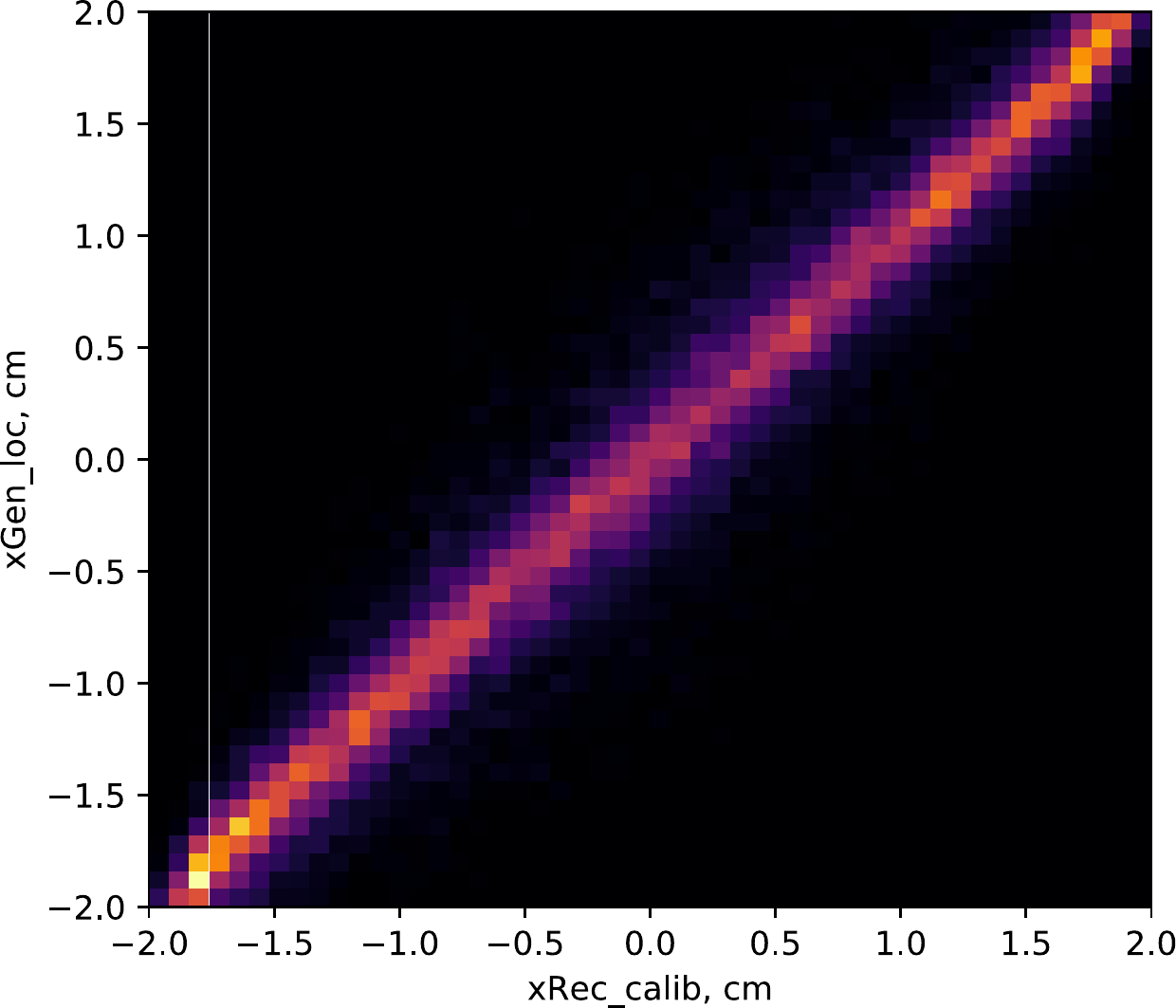}
    \end{minipage}%
\caption{\label{fig:S-curve_calibration} S-curve calibration results for inner modules and local coordinate \textit{x} using parametric approach (left) and XGBoost regressor (right).}
\end{figure}

For each photon from the signal sample, we find the closest track in \textsc{Geant4} simulated data.\footnote{The track affinity is based on distance in \textit{px, py, pz} space.
For quick nearest-neighbour lookup the 3 dimensional kd-tree was created using the package cKDTree from SciPy\cite{scipy}.}
% At this point, the clusterisation procedure begins.
The calorimeter cell, in which the signal produces the hit is required to be surrounded by two layers of cells of the same type.
Thus, a matrix of 5x5 cells of the same type is obtained.\footnote{It is assumed that the use of adjacent modules of the same type is also suitable for the borders between calorimeter regions. In this case, the adjacent module of another type on the other side of the border will be surrounded by modules of its own type. The contribution of responses of modules of different type will be compensated.}
% The cell We shift the search window of 5x5 modules in steps of 1 module and each time we search for the cell with the highest energy deposit.
We suppose that most of the clusters of the signal sample do not exceed the size of such a matrix. Inside this matrix, the cell with the highest energy deposit is searched.
% Its energy, as well as the energies of all the adjacent cells in two layers (25 cells in total), are recorded.
The barycentre of the cluster is the reconstructed position of the photon which released energy in the calorimeter.
The dependence of each of the local coordinates of the signal cluster barycentre on the corresponding true coordinate of the hit position we call the S-curve due to its distinctive shape.
The S-curves of both \textit{x} and \textit{y} coordinates for the inner region are displayed in Figure~\ref{fig:inner_0_nPV_s_curve_x_and_y}. The difference between the S-curve and the straight line characterises the quality of spatial reconstruction. Several approaches were tested to calibrate the S-curve (hit position reconstruction): the parametric approach and the machine learning approach using XGBoost regressor\cite{xgboost}. The results of the calibration for these approaches are shown in Figure~\ref{fig:S-curve_calibration} and in Table~\ref{tab:calibration_results}.

% It is important to take into account local granularity because we require this approach to operate within arbitrary arrangement of modules of different types.

As a metric of spatial resolution we use RMSE of the difference between true and reconstructed local coordinate (independent of \textit{x} and \textit{y}) of the hit. The observed difference in the metric, based on local coordinates \textit{x} and \textit{y} was found to be negligible. Therefore, all the results exploited in this metric are presented for local coordinate \textit{x}.

For the parametric approach, the reconstructed local coordinate \textit{x} is represented as ${a \cdot arcsinh(b \cdot x)}$. The parameters for the calibration were found using random search\cite{random} with 1000$^2$ points in the range (0.01, 100) for each parameter. The best parameters obtained using a parametric approach for the inner section are: \textit{a} = 1.15, \textit{b} = 2.07.

The selected machine learning approach was based on the extreme gradient boosting (XGBoost), and among its hyperparameters \textit{colsample\_bytree, gamma, max\_depth} and \textit{min\_child\_weight} were selected, which are typical for such a problem. These hyperparameters were optimised using BayesSearchCV\footnote{A package from scikit-optimize\cite{skopt}.}, within the ranges (1, 20), (1, 10), (0.1, 0.9) and (0.3, 0.7), respectively. The chosen parameters \textit{colsample\_bytree} = 0.7, \textit{gamma} = 0.1, \textit{max\_depth} = 20, \textit{min\_child\_weight} = 10 (the set of values is for the inner section) provide best results without overtraining being observed. As a result, we used 5-fold cross-validation and trained the regressor using 30\% events of the sample.

% \usepackage{booktabs}

% \begin{table}[!htb]
% \begin{center}
% \begin{tabular}{@{}llll@{}}

% \toprule
% \textbf{ECAL region}  & Inner  & Middle & Outer  \\ \midrule
% \textbf{Calibration} &&&\\
% Uncalibrated & 0.5471 & 0.9262 & 2.2903 \\
% Parametric   & 0.2734 & 0.5018 & 1.0365 \\
% XGBoost      & 0.1219 & 0.284  & 0.7573 \\ \bottomrule
% \end{tabular}
% \end{center}
% \caption{\label{tab:calibration_results} Spatial reconstruction results (RMSE, cm) of local coordinate \textit{x} for uncalibrated data, and for the data calibrated using parametric approach and XGBoost regressor.}
% \end{table}

\begin{table}[!htb]
\begin{center}
\begin{tabular}{@{}llll@{}}

\toprule
\textbf{ECAL region}  & Inner  & Middle & Outer  \\ \midrule
Uncalibrated & 0.860 $\pm$ 0.007 & 1.413 $\pm$ 0.017 & 3.795 $\pm$ 0.030 \\
\textbf{Calibration} &&&\\
Parametric   & 0.522 $\pm$ 0.009 & 0.964 $\pm$ 0.012 & 2.780 $\pm$ 0.025 \\
XGBoost      & 0.117 $\pm$ 0.002 & 0.267 $\pm$ 0.004 & 0.731 $\pm$ 0.002 \\ \bottomrule
\end{tabular}
\end{center}
\caption{\label{tab:calibration_results} Spatial reconstruction results (RMSE, cm) of local coordinate \textit{x} for uncalibrated data, and for the data calibrated using parametric approach and XGBoost regressor.}
\end{table}

% Alternative table design
% \begin{table}[hbt]
% \begin{tabular}{@{}llll@{}}
% \toprule
% \backslashbox[48mm]{Calibration}{ECAL region}
% &\makebox[3em]{Inner}&\makebox[3em]{Middle}&\makebox[3em]{Outer} \\ \midrule
% Uncalibrated & 0.5471 & 0.9262 & 2.2903 \\
% Parametric   & 0.2734 & 0.5018 & 1.0365 \\
% XGBoost      & 0.1219 & 0.284  & 0.7573 \\ \bottomrule
% \end{tabular}
% \end{table}

%The dataset is aused as training sample for LHCb CaloGAN[ADD LINK].\\ 
% To reconstruct ECAL clusters by calculating center-of-mass of energy deposits of given cluster. % Spatial Reconstruction
\section{Pile-up Mitigation with Timing}
Among the requirements for the LHCb Phase 2 Upgrade ECAL, is the ability to measure the time of arrival of the photon or electron with an accuracy of few tenths of a picosecond. The difficulty is that the time-of-arrival properties are hard to reproduce in an accurate simulation. We used test beam results to evaluate important simulation parameters, as well as calibrate simulation on points measured during the test beam. The data was obtained from the electron test beam\footnote{Data obtained from the 30 GeV electron beam @\textsc{DESY} for LHCb electromagnetic calorimeter module} and consists of 7848 signals, where each signal contain 1024 impulse measurements sampled with a step of 200 picoseconds (5~GHz) involving the reference time of the signal. The time of arrival has been studied in two cases: single signal and two overlapped signals. The goal of using machine learning algorithms is to evaluate limitations of the possible physics performance driven by test beam data.

\subsection{Single signal} \label{sec:single_signal}
In the case of a single signal, the machine learning algorithms were used to evaluate its reference time.
It has been investigated how accurately the reference time can be reconstructed depending on the sampling rates. The second case
is based on reference time reconstruction, in the presence of the second signal. Additionally, machine learning was applied for the classification in which cases the data contain one signal, or multiple ones.

% Overall, there are two major cases that we are going to consider. The first one is a case of a unique
% signal. There we would like to apply machine learning algorithms in order to evaluate the reference
% time of signals. Moreover, we explored how precisely we can recover the reference time depending
% on different sampling rates.

% Additionally, machine learning was applied in order to
% classify whether we have one signal data or we have multiple signals.
% To clarify, the aim of applying machine learning to this problem was not beating the score of
% particular baseline algorithm. Oppositely, The goal is to use ML to extract the maximum of
% available in data information and evaluate limitations of the possible physics performance
% that are driven by behaviours of actual, test beam data. Furthermore, we would like to explore how
% the quality of algorithms depends on the stability of the data.\\

% The data was obtained from the electron test beam, overall, the dataset consists of 7848 signals,
% where each signal is 1024 impulse measurements, sampled with a step 200 picoseconds (5~GHz) +
% the reference time of a signal.\\

\begin{figure} [hbt]
\centering
    \begin{minipage}{.5\textwidth}
        \centering
        \includegraphics[width=\linewidth]{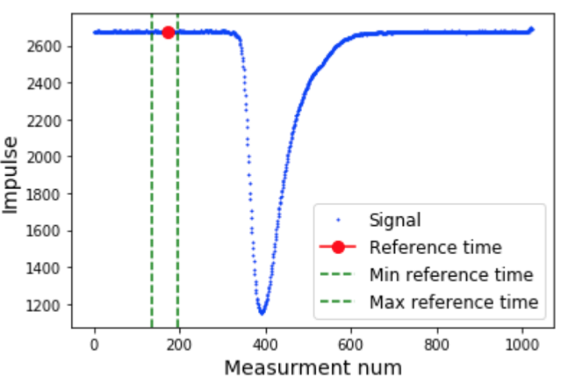}
    \end{minipage}%
\centering
    \begin{minipage}{.5\textwidth}
        \centering
        \includegraphics[width=\linewidth]{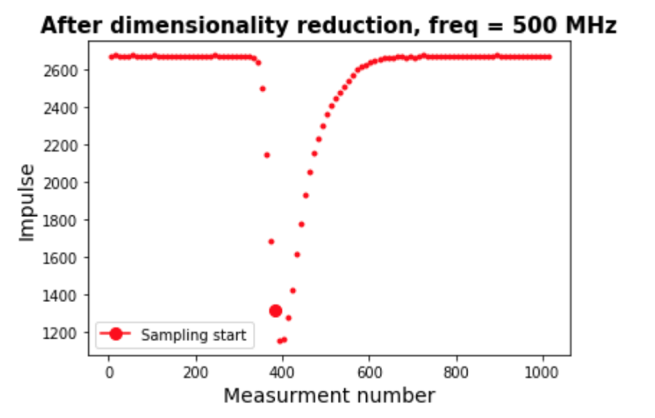}
    \end{minipage}%
    \caption{\label{fig:signal_sampling} Signal sampling before (left) and after dimensionality reduction (right).}
\end{figure}

The raw signal obtained during the test beam, is presented in Figure~\ref{fig:signal_sampling} (left). Artificial re-sampling allows us to reduce data while retaining the signal shape as shown in Figure~\ref{fig:signal_sampling} (right).

% We were strongly interested in the way the quality of recovering depends on a
% sampling frequency. In order to explore the dependency, the data transformation had to be done. We
% did it in a straightforward way.

The procedure of test beam data analysis starts with the choosing of an initial sampling point. Afterwards, given a
frequency k, each k-th point was sampled towards both lower and higher frequencies.\\

\begin{figure} [hbt]
\centering
\includegraphics[width=0.6\linewidth]{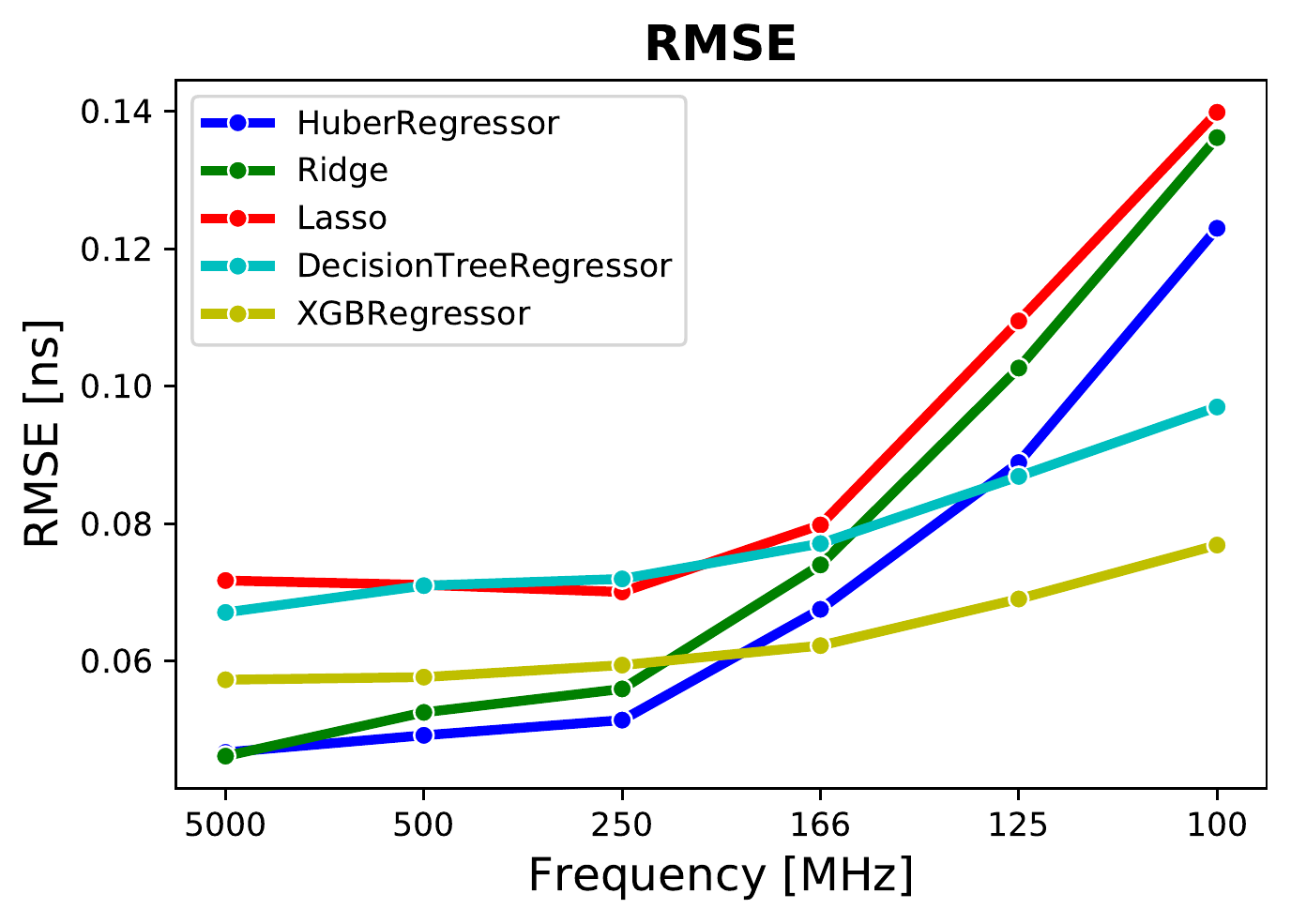}
\caption{\label{fig:models_comparison_2} Models comparison for time difference RMSE (ns) w.r.t. sampling frequency (MHz).}
\end{figure}

The reference time was predicted using 5 different models. All the regressors have been tuned using Bayesian optimisation.
Feature engineering for the selected regressors did not show any significant improvement in the results.
% The key reason of doing it was that we wanted to verify the consistency of our results.
Time difference RMSE was chosen as a loss function.
% Additionally, all the data was preprocessed by min-max scaling.\\
The comparison of 5 selected models is shown in Figure~\ref{fig:models_comparison_2}.
Different regressors demonstrate similar results. One can see that the sampling frequency can be reduced from 5~GHz to 250~MHz without considerable changes of time reconstruction. Subsequent results were obtained using the XGBoost regressor\cite{xgboost}.
% More detailed XGBoost regressor output vs. Sampling frequency is displayed in Figure~\ref{fig:models_comparison_2} (right).
% Feature engineering did not show any significant improvement in the results.
% As the line graph reveals, we can reduce sampling frequency by 25 times without any considerably
% meaningful changes in the quality of time recovery.
% As well, some experiments with feature
% engineering were done, such as an addition of polynomial features, that didn’t involve any growth
% in the metric.\\

\subsection{Overlapping signals}
High pile-up conditions imply overlap of signals from different vertices. Timing information can be used to mitigate pile-up. Hereinafter, we consider signal processing at the readout for an individual calorimeter cell.

\begin{figure} [hbt]
\centering
    \begin{minipage}{.55\textwidth}
        \centering
        \includegraphics[width=\linewidth]{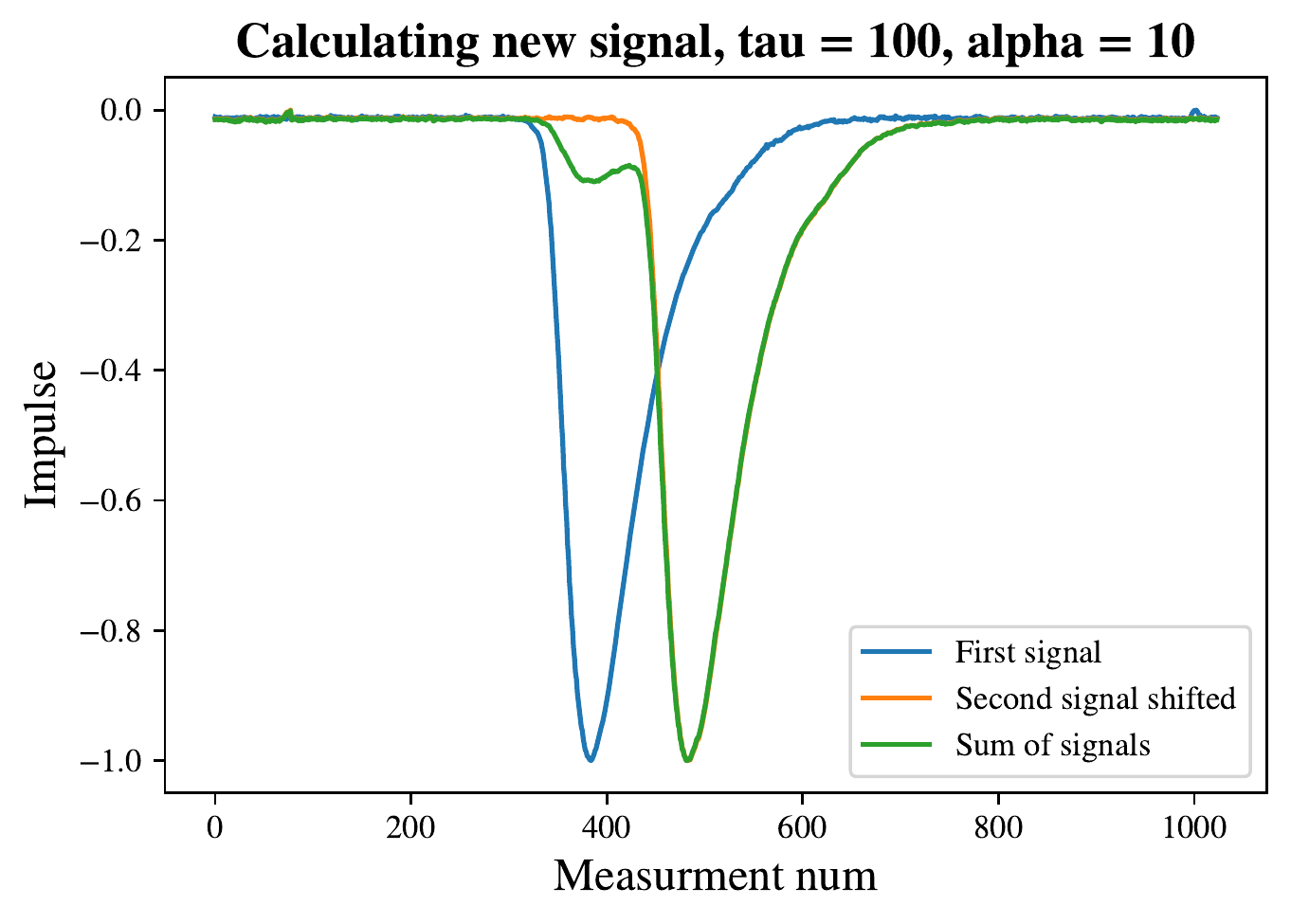}
    \end{minipage}%
\centering
    \begin{minipage}{.45\textwidth}
        \centering
        \includegraphics[width=\linewidth]{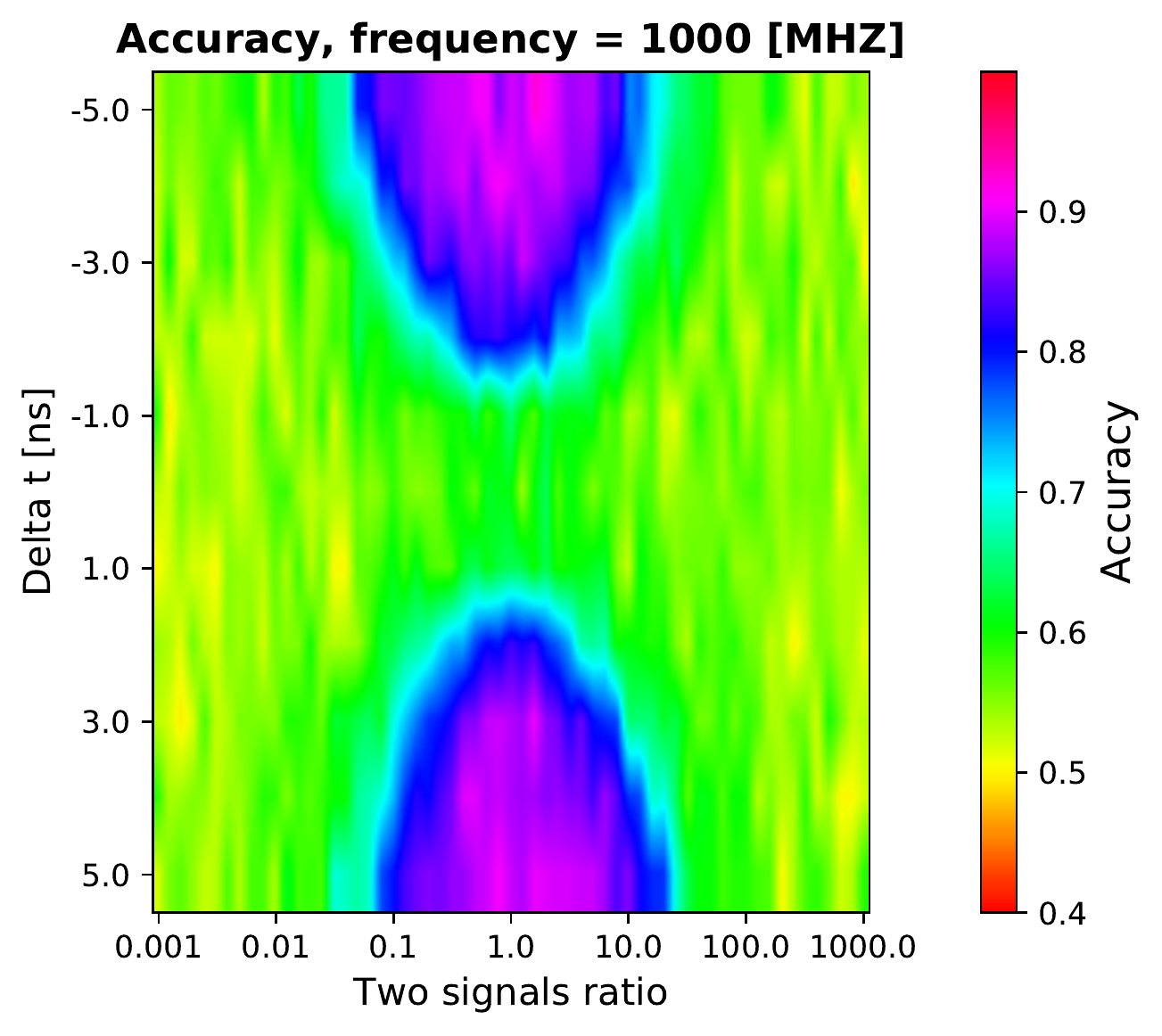}
    \end{minipage}%
    \caption{\label{fig:overlapping_signals_1} Generated and initial signals (left). The 2-dimensional distribution of arrival time difference and signal ratio of two signals for sampling frequency 1000 MHz (right).}
\end{figure}

Since we didn't have samples with multiple signals, we generated it under given amplitude ratio and time shift. The resulted signal can be parameterised as $S_{\mathrm{res}} = S_1 + \alpha\cdot{}S_2^{\mathrm{shifted}}$, where $S_1$ and $S_2$ are different signals obtained from the test beam data in the same way as described in Section~\ref{sec:single_signal}. The reference time of the signal which amplitude is greater is assigned to the reference time of the resulting signal. Figure~\ref{fig:overlapping_signals_1} (left) shows the generated signal. By aggregation of the arrival time difference ($\tau$) and ratio ($\alpha$) of two signals one can produce 2-dimensional distributions over these variables as displayed in Figure~\ref{fig:overlapping_signals_1} (right). It demonstrates the accuracy of our predictions on these two parameters for two different datasets.

\subsection{Reference time prediction in the presence of the second signal}
The final model for time difference/amplitude separation of two signals employs an ensemble of models of KNeighborsClassifier, DecisionTreeClassifier and RandomForestClassifier.\footnote{These models are available as packages from Scikit-Optimize\cite{skopt}.}
The models were tuned using Bayesian optimisation\footnote{Using Hyperopt\cite{hyperopt}} separately for each sampling frequency.
A variety of strategies we tried for probabilistic class estimation, such as ‘linear’,
‘harmonic‘, ‘geometric‘ and ‘rank averaging‘. ‘Harmonic‘ resulted as the best one according to the cross-validation.

\begin{figure} [hbt]
\centering
\includegraphics[width=0.6\linewidth]{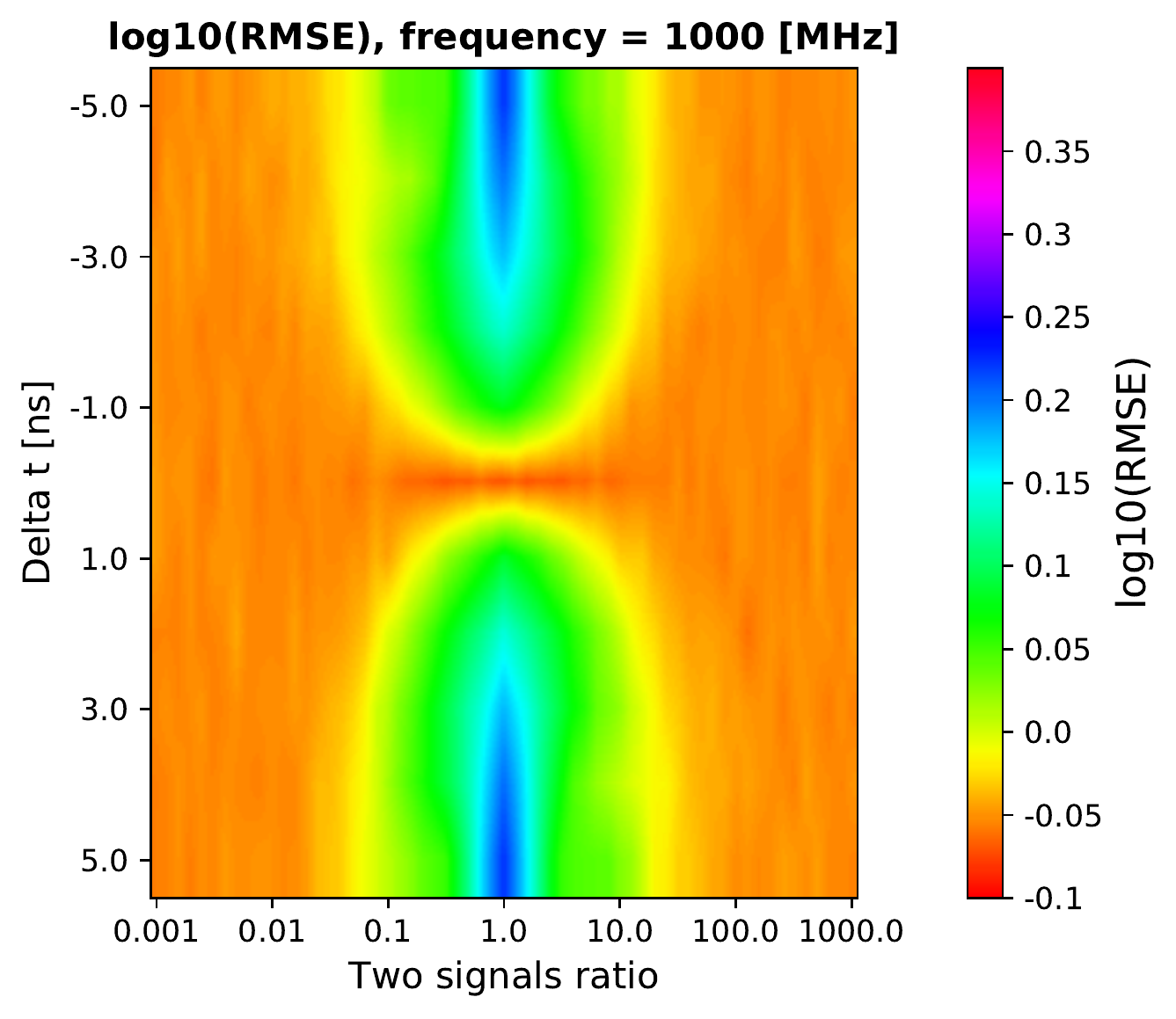}
\caption{\label{fig:overlapping_signals_final} The 2-dimensional distribution of arrival time difference and signal ratio of two signals for sampling frequency 1000 MHz.}
\end{figure}

Figure~\ref{fig:overlapping_signals_final} demonstrates the ability of the final model to evaluate time reconstruction in the presence of the second signal. 

% The following classifiers were chosen as basic models: KneighborsClassifier,
% DecisionTreeClassifier, RandomForestClassifier. Nevertheless, the experiments were held with
% other classifiers as well, such as GaussianNB, PassiveAggressiveClassifier and XgboostClassifier
% as well. However, the usage of GaussianNB and PassiveAggressiveClassifier didn’t result in any
% score improvement.
% While, the usage of Xgboost affected the speed of inference greatly, while,
% didn’t cause a significant metric improvement (accuracy).
% Furthemore, the models were tuned using Bayesian optimisation (hyperopt) separately for each sampling frequency.\\

% Additionally, we wanted to explore the dependency of the accuracy of our predictions on two
% parameters: tau and alpha for two different datasets. This dependency is illustrated on the above
% images.\\

% Finally, we wanted to learn to extract the reference time of unique signals that resulted in a multiple
% signal. Ideally, it would be useful to recover the reference time of both signals. Unfortunately, our
% models have shown that the quality of a reconstruction for a signal with a smaller amplitude is
% incredibly low. That being said, we decided to follow up only with a signal that has a greater
% amplitude.\\

% When it comes to a model, XGBRegressor was used, while the optimal hyperparameters were tuned
% using Bayesian optimisation separately for each sampling frequency.\\ % Pile-up Mitigation with Timing

\section{Conclusion}

The proposed approaches of spatial reconstruction and time of arrival properties of the LHCb electromagnetic calorimeter illustrate the idea of automation of the detector development and its optimisation cycle. The results obtained, using the discussed approaches, addresses physics performance aspects which are known to be computationally expensive for accurate simulation. The spatial reconstruction using both parametric and machine learning approaches was obtained. The performance of the selected XGBoost configuration surpass those of the parametric approach for each of calorimeter regions. For the pile-up mitigation, the machine learning approach demonstrates the ability to evaluate time reconstruction in the presence of the second signal. 

\section{Acknowledgements}
The research leading to these results has received funding from Russian Science Foundation under grant agreement n$^{\circ}$~19-71-30020.

\end{document}